\documentclass[twocolumn,prl,showpacs]{revtex4}
\usepackage{amsmath,eucal}
\usepackage{graphicx}
\usepackage{pifont}
\usepackage{verbatim}

\begin{document}
\title{Controlling quantum transport through a single molecule} 
\author{D. M. Cardamone, C. A. Stafford, S. Mazumdar}
\affiliation{Department of Physics, University of Arizona, 1118 E. 4th Street,
Tucson, AZ 85721}

\date{\today}

\begin{abstract}

We investigate multi-terminal quantum transport through single monocyclic aromatic annulene molecules,
and their derivatives, using the nonequilibrium Green function approach in the self-consistent
Hartree-Fock approximation.  A new device concept, the Quantum Interference Effect
Transistor (QuIET) is proposed, exploiting perfect destructive interference stemming from
molecular symmetry, and controlling current flow by introducing decoherence and/or elastic scattering
that break the symmetry.  This approach overcomes the fundamental problems of power dissipation and
environmental sensitivity that beset 
many nanoscale device proposals.

\end{abstract}


\maketitle

From the vacuum tube to the modern CMOS transistor, devices which control 
the flow of electrical current
by modulating an electron energy barrier are ubiquitous in electronics.
In this paradigm, 
a minimum energy of $k_B T$ must be dissipated 
to switch the current ``on'' and ``off,'' 
necessitating incredible power dissipation
at device densities approaching the atomic limit \cite{roadmap}. 
A possible alternative 
is to control electron flow using quantum interference \cite{sautet88,sols89,baer02,stadler03}.  
In mesoscopic devices, quantum interference is typically tuned via the Aharanov-Bohm effect \cite{washburn86};
however, in nanoscale conductors such as single molecules, 
this is impractical due to the enormous magnetic fields required to produce a phase shift of order one radian.
Similarly, a device based on an electrostatic phase shift \cite{sols89,baer02} would, in small molecules, require voltages 
incompatible with structural stability.
We propose a solution exploiting perfect destructive interference stemming from molecular symmetry,
and controlling quantum transport by introducing decoherence or scattering from a third lead.

\begin{figure}[b]
\includegraphics[width=0.95\columnwidth,keepaspectratio=true]{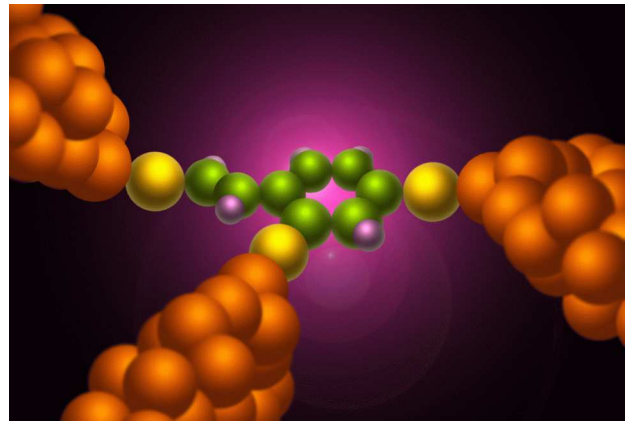}
\caption{
Artist's conception of a Quantum Interference Effect Transistor
(QuIET).  The colored spheres represent individual carbon (green),
hydrogen (purple), and sulfur (yellow) atoms, while the three gold
structures represent the metallic contacts.  A voltage applied to the
leftmost contact regulates the flow of current between the other two.
}
\label{fig:QuIET_3d}
\end{figure}

As daunting as the fundamental problem of the switching mechanism, is the practical problem of nanofabrication \cite{roadmap}.
In this respect, single molecules have a distinct advantage over other types of nanostructures, in that large numbers of identical devices
can be readily synthesized.  Single-molecule devices with two leads have been fabricated by a number of techniques
\cite{nitzan03}.  Our transistor
requires a third terminal coupled locally to the molecule,
capacitively or via tunneling (see Fig.~\ref{fig:QuIET_3d}).  
To date, only global gating of single-molecule devices has been achieved \cite{nitzan03}; recently, however,
there has been significant progress toward a locally coupled third terminal \cite{piva05}.

This Letter reports the results of our recent theoretical investigations into the
use of interference effects to create molecular transistors, leading to a
new device concept, which we call the Quantum Interference Effect Transistor (QuIET). 
We demonstrate that for all monocyclic aromatic annulenes, particular two-terminal configurations exist 
in which destructive interference blocks current flow, and that transistor
behavior can be achieved by supplying tunable decoherence or scattering at a third
site. 
We also propose a realistic model for
introducing scattering in a controllable way, using an alkene chain
of arbitrary length ({\it cf}.\ Fig.~\ref{fig:QuIET_3d}). Finally, we present 
nonequilibrium Green function (NEGF) calculations within the 
self-consistent Hartree-Fock approximation, indicating that the QuIET functions at room temperature
with a current-voltage characteristic strikingly
similar to macroscale transistors.

The Hamiltonian of the system can be written as the sum of three terms:
$H=H_{mol}+H_l+H_{tun}$. The first is the 
$\pi$-electron molecular Hamiltonian
\begin{multline}
\label{Hm}
H_{mol}=\sum_{n\sigma}\varepsilon_n d_{n\sigma}^\dagger d_{n\sigma}
-\sum_{\langle nm\rangle\sigma}\left(t_{nm}d_{n\sigma}^\dagger
  d_{m\sigma}+\mathrm{H.c.}\right)\\ 
\mbox{ }+\sum_{nm}\frac{U_{nm}}{2}Q_nQ_m,
\end{multline}
where $d^\dagger_{n\sigma}$ creates an electron of spin $\sigma$ in the $\pi$-orbital of the $n$th carbon atom,
$\varepsilon_n$ are the orbital energies, and $\scriptstyle \langle \mbox{ } \rangle$ indicates a sum over nearest neighbors. 
The tight-binding hopping
matrix elements $t_{nm}=2.2\mbox{eV}$, $2.6\mbox{eV}$, or $2.4\mbox{eV}$ for 
orbitals
connected by a single bond, double bond, or within an aromatic ring, respectively.
The final term of Eq.\ (\ref{Hm}) contains intra- and intersite
Coulomb interactions, as well as the electrostatic effects of the
leads. The interaction energies are given by 
the Ohno parameterization \cite{ohno64,chandross97}:
\begin{equation}
\label{interactions}
U_{nm}=\frac{11.13\mathrm{eV}}{\sqrt{1+.6117\left(R_{nm}/\mathrm{\AA}\right)^2}},
\end{equation}
where $R_{nm}$ is the distance between orbitals $n$ and $m$.
$Q_n=\sum_\sigma d_{n\sigma}^\dagger
d_{n\sigma}-\sum_\alpha C_{n\alpha}\mathcal{V}_\alpha/e -1$ is an
effective charge operator \cite{stafford98} for orbital $n$, where
the second term represents a polarization charge.
Here $C_{n\alpha}$ is the capacitance between orbital $n$ 
and lead $\alpha$, chosen consistent 
with the interaction energies of Eq.\
(\ref{interactions}) and the geometry of the device, and $\mathcal{V}_\alpha$ is the voltage on lead
$\alpha$. $e$ is the magnitude of the electron charge.

\begin{figure}[t]
    \includegraphics[trim=3cm 1cm 3cm 2cm,width=.667\columnwidth,keepaspectratio=true,clip=true]{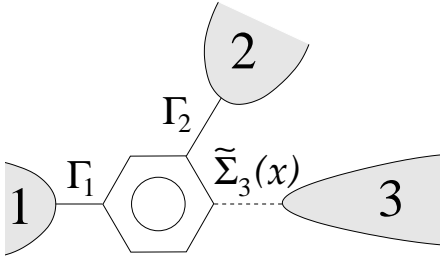}
  \caption{Schematic diagram of a 
  QuIET based on benzene. Here $\Gamma_{1,2}$ are the coupling strengths of metallic leads 1 and 2, connected in the meta orientation, 
  to the corresponding $\pi$-orbitals of benzene.
  $\tilde{\Sigma}_3$, determined by a control variable $x$, is the retarded self-energy induced by lead 3.
  The real part of $\tilde{\Sigma}_3$ introduces
  elastic scattering, while the imaginary part introduces decoherence. 
  }
  \label{QuIET}
\end{figure}

Each metal lead $\alpha$ possesses a continuum of states, and their total Hamiltonian is
\begin{equation}
H_l=\sum_{\alpha}\sum_{\substack{k\in\alpha\\ \sigma}}\epsilon_k c_{k\sigma}^\dagger c_{k\sigma},
\end{equation}
where $\epsilon_k$ are the energies of the single-particle levels in the leads, 
and $c^\dagger_{k\sigma}$ is an electron creation operator.
Tunneling between molecule and leads is provided by the final term of the
Hamiltonian,
\begin{equation}
H_{tun}=\sum_{\langle n\alpha\rangle}\sum_{\substack{k\in\alpha\\ \sigma}}\left(V_{nk}d_{n\sigma}^\dagger
  c_{k\sigma}+\mathrm{H.c.}\right),
\end{equation}
where $V_{nk}$ are the tunneling matrix elements from a level $k$ within
lead $\alpha$ to the nearby site $n$. Coupling of the leads to the molecule via
molecular chains, as may be desirable for fabrication purposes, can be
included in the effective $V_{nk}$, as can the effect of substituents (e.g., thiol groups)
used to bond the leads to the molecule \cite{tian98,nitzan01}. 

We use the 
NEGF approach \cite{jauho94,datta95} to describe transport in this open quantum system.
Given the retarded Green function of the isolated molecular system $G_{mol}(E)=\left(E-H_{mol}+i0^+\right)^{-1}$,
Dyson's equation 
\begin{equation}
\label{Dyson}
G(E)=\left[G_{mol}^{-1}(E)-\Sigma(E)\right]^{-1}
\end{equation}
gives the Green function of the full system.
The QuIET is intended for use at room temperature and above, 
and operates in a voltage regime where there are no unpaired electrons in the molecule.
Thus lead-lead and lead-molecule correlations, such as the Kondo effect, do not play an important role.
Electron-electron interactions may therefore be included via the self-consistent Hartree-Fock
method \cite{marder00}. $H_{mol}$ is replaced by the corresponding 
mean-field Hamiltonian $H_{mol}^{HF}$, which is quadratic in electron creation and annihilation operators,
and contains long-range hopping.  Within mean-field theory, the retarded self-energy due to the leads is 
\begin{equation}
\label{self_energy}
\Sigma_{n\sigma,m\sigma'}(E)=-\frac{i}{2}\delta_{nm}\delta_{\sigma\sigma'}\sum_{\langle a\alpha\rangle}\Gamma_\alpha(E)\delta_{na},
\end{equation}
where $\Gamma_\alpha(E)=2\pi\sum_{k\in\alpha}|V_{nk}|^2\delta\left(E-\epsilon_{ k}\right)$
is the Fermi's Golden Rule tunneling width.
As a result, the molecular density of states 
changes from a discrete spectrum of delta functions to a continuous, width-broadened distribution. 
We take the broad-band limit \cite{jauho94}, treating $\Gamma_\alpha$ as constants characterizing the coupling of the leads to the molecule. 
Typical estimates \cite{nitzan01} using the  method of Ref.\ \cite{mujica94} yield
$\Gamma_\alpha \lesssim 0.5\mbox{eV}$, but values as large as 1eV have been suggested \cite{tian98}. 

The effective hopping and orbital energies in 
$H_{mol}^{HF}$ depend on the equal-time correlation functions, which are found in the NEGF approach 
to be
\begin{equation}
\label{correlations}
\langle d_{n\sigma}^\dagger
d_{m\sigma}\rangle=\sum_{\langle a\alpha\rangle}\frac{\Gamma_\alpha}{2\pi} \int_{-\infty}^\infty \!\!\! dE \,
G_{n\sigma,a\sigma}(E)G^*_{a\sigma,m\sigma}(E)f_\alpha(E),
\end{equation}
where $f_\alpha(E)=\{1+\exp[(E-\mu_\alpha)/k_B T]\}^{-1}$ is the Fermi function for lead $\alpha$. 
Finally, the Green function is determined by iterating the self-consistent loop, Eqs.\ (\ref{Dyson})--(\ref{correlations}).

The current in lead $\alpha$ is given by the multi-terminal current formula \cite{buttiker86}
\begin{equation}
\label{l-b}
I_\alpha=\frac{2e}{h}\sum_\beta\int_{-\infty}^\infty
dE\;T_{\beta\alpha}(E)\left[f_\beta(E)-f_\alpha(E)\right],
\end{equation}
where $T_{\beta\alpha}(E)=\Gamma_\beta\Gamma_\alpha|G_{ba}(E)|^2$
is the tranmission probability \cite{datta95} from lead $\alpha$ to lead $\beta$, 
and $a\, (b)$ is the orbital coupled to lead $\alpha\, (\beta)$.

The QuIET exploits quantum interference stemming from the symmetry of monocyclic aromatic annulenes, such as benzene. 
Quantum transport through single benzene molecules with two metallic leads connected in the {\it para} orientation has
been the subject of extensive experimental and theoretical investigation \cite{nitzan03}; however, a QuIET based on 
benzene requires the source (1) and drain (2) to be connected in a {\it meta} orientation, as illustrated in Fig.~\ref{QuIET}.
An electron propagating between leads 1 and 2 takes all possible paths within the molecule.
In the absence of a third lead ($\tilde{\Sigma}_3=0$), these paths all lie within the benzene ring.
In linear response, and with no net charge transfer between molecule and leads,
each electron injected into the molecule has momentum given by its Fermi wavenumber,
$k_F=\pi/2d$, where $d=1.397\mathrm{\AA}$ is the intersite spacing of
benzene. The phase difference between the two most direct
paths through the ring is $\pi$, and they interfere destructively.
Similarly, all of the paths through the ring 
cancel exactly in a pairwise fashion. 
It is a consequence of Luttinger's Theorem \cite{luttinger} that this coherent suppression of current is not altered by electron-electron
interactions. 

\begin{figure}[t]
  \setlength\unitlength{.333\columnwidth}
  \begin{picture}(3,3)
    \put(0,2){\includegraphics[width=\columnwidth,height=\unitlength]{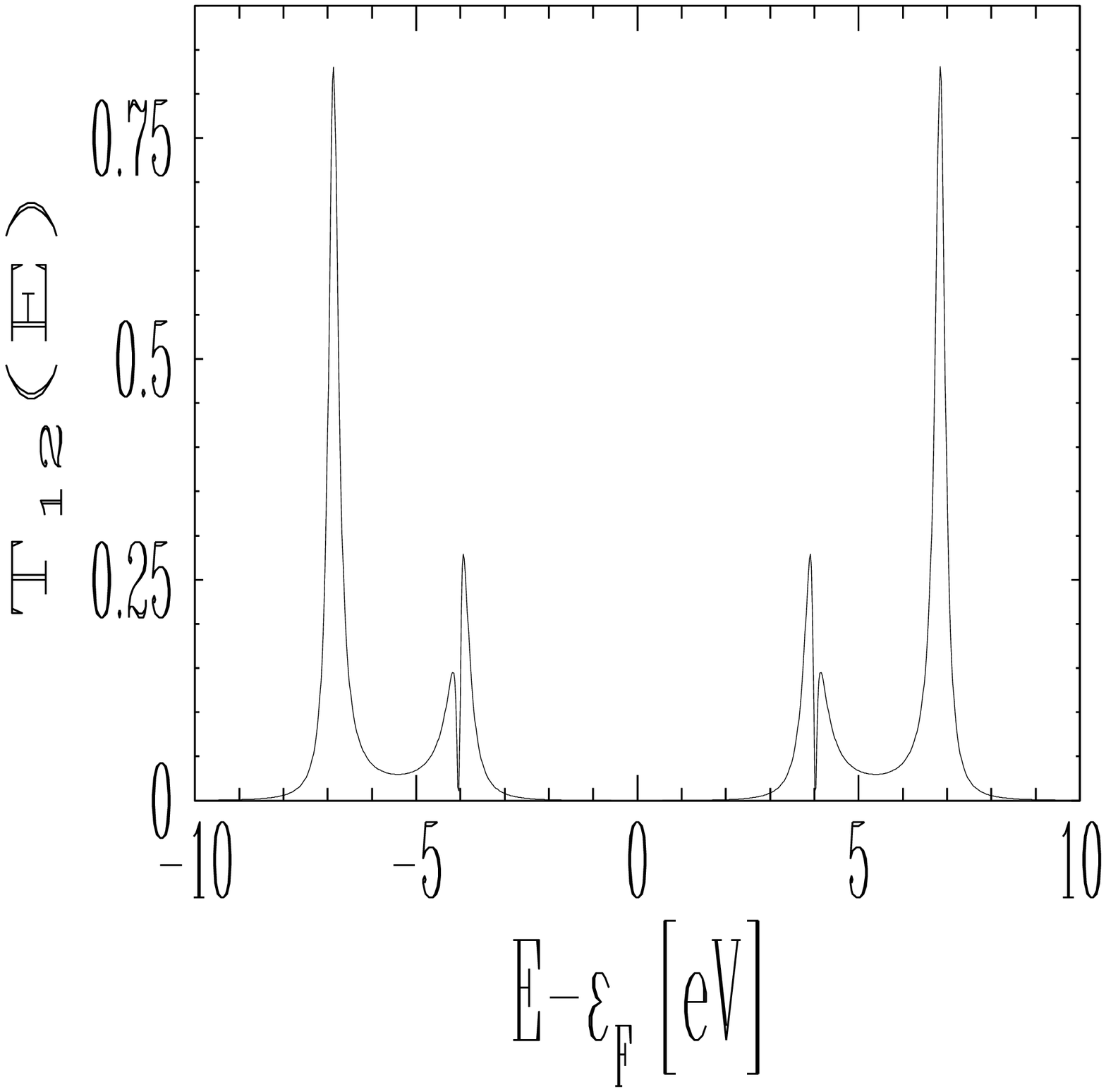}}
    \put(0,1){\includegraphics[width=\columnwidth,height=\unitlength]{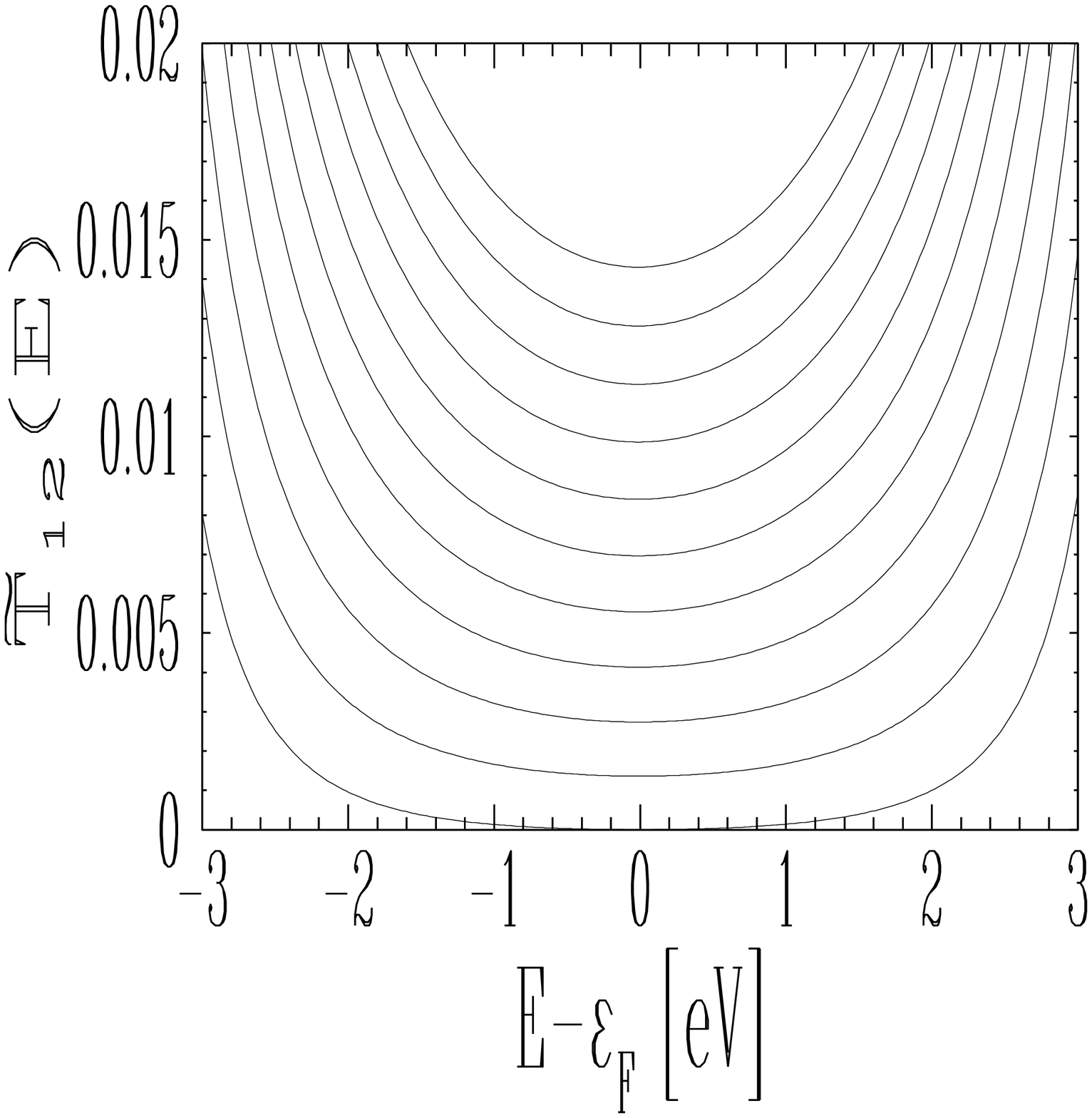}}
    \put(0,0){\includegraphics[width=\columnwidth,height=\unitlength]{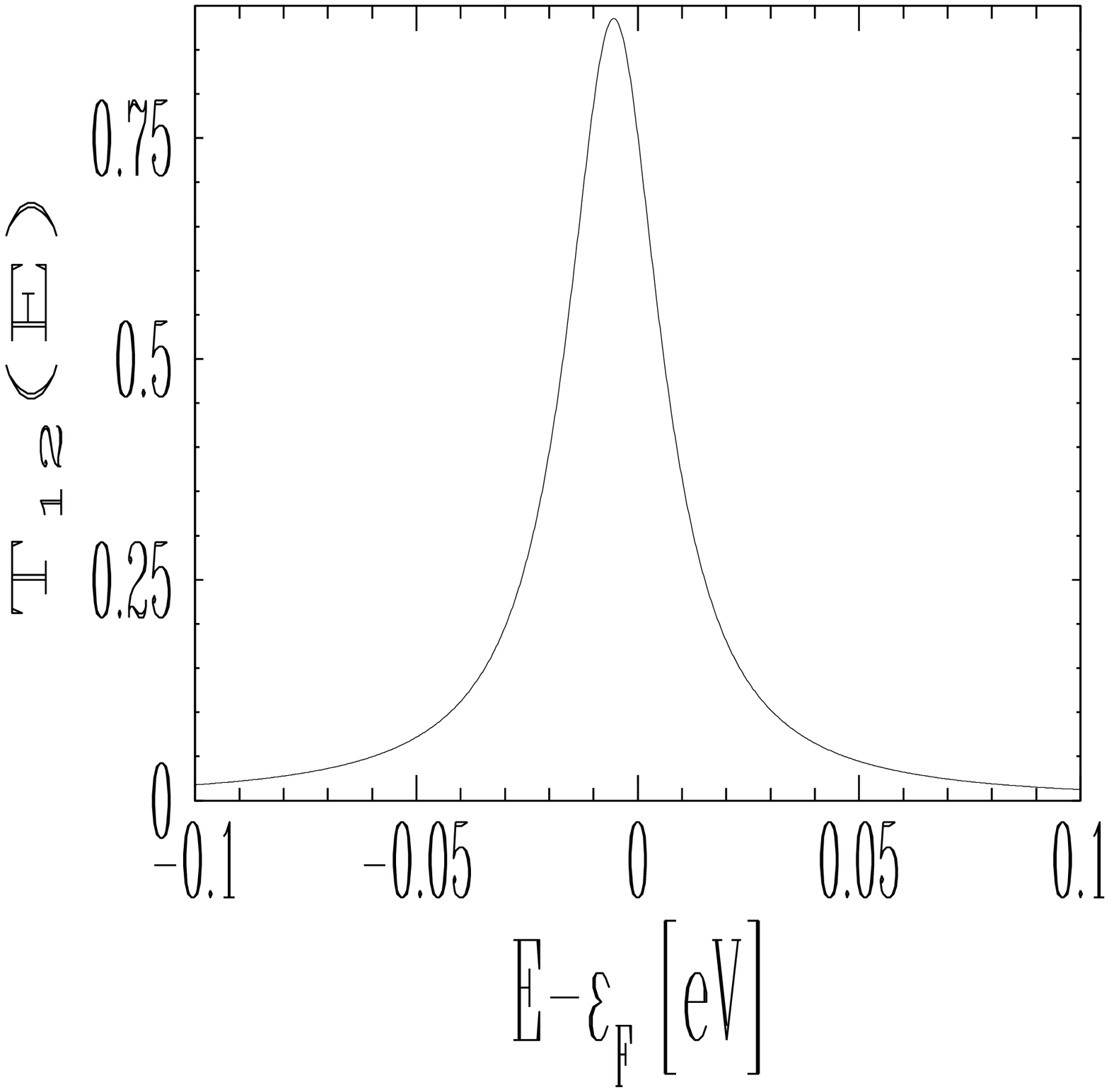}}
    \put(0,2.95){(a)}
    \put(0,1.95){(b)}
    \put(0,.95){(c)}
  \end{picture}
  \caption{Effective transmission probability $\tilde{T}_{12}$
  of the device shown in Fig.~\ref{QuIET}, at room temperature,
  with $\Gamma_1=1.2\mathrm{eV}$ and
  $\Gamma_2=.48\mathrm{eV}$: (a) $\tilde{\Sigma}_3=0$; 
  (b) $\tilde{\Sigma}_3=-i\Gamma_3/2$, where
  $\Gamma_3=0$ in the lowest curve and increases by .24eV in each
  successive one; (c) $\tilde{\Sigma}_3$ is given by Eq.\ (\ref{realsigma}) with a single resonance, 
  $\varepsilon_\nu=\varepsilon_F$ and $t_\nu=1\mathrm{eV}$.}
  \label{suppression}
\end{figure}


Figure \ref{suppression}a shows the transmission probability $T_{12}$ of the device shown in Fig.~\ref{QuIET}
for $\tilde{\Sigma}_3(E)=0$, illustrating
the total current suppression at the Fermi energy (see also Fig.~\ref{suppression}b, lowest curve).
This suppression can be lifted by introducing decoherence or elastic scattering that break the molecular symmetry. 
Figures \ref{suppression}b and c illustrate the effect of attaching a third lead to the molecule as shown in Fig.~\ref{QuIET}, 
introducing a complex self-energy $\tilde{\Sigma}_3(E)$ on the $\pi$-orbital adjacent to that connected to lead 2.

An imaginary self-energy $\tilde{\Sigma}_3=-i\Gamma_3/2$ corresponds to coupling a third metallic lead directly to the
benzene molecule. 
If the third lead functions as an infinite-impedance voltage probe, the effective two-terminal transmission is \cite{buttiker88}
\begin{equation}
\tilde{T}_{12}=T_{12}+\frac{T_{13}T_{32}}{T_{13}+T_{32}}.
\end{equation}
The third lead introduces decoherence \cite{buttiker88} and additional paths that are not cancelled, thus allowing current to flow,
as shown in Fig.~\ref{suppression}b.
As a proof of principle, a QuIET could be constructed using a scanning
tunneling microscope tip as the third lead, with tunneling coupling $\Gamma_3(x)$
to the appropriate $\pi$-orbital of the benzene ring, 
the control variable $x$ being the piezo-voltage controlling the tip-molecule distance.

By contrast,
a real self-energy $\tilde{\Sigma}_3$ introduces elastic scattering, which can also break the molecular symmetry.
This can be achieved by attaching a second molecule to the benzene ring, for example an alkene chain ({\it cf}.\ Fig.~\ref{fig:QuIET_3d}).
The retarded self-energy due to the presence of a second molecule is 
\begin{equation}
\label{realsigma}
\tilde{\Sigma}_3(E)=\sum_\nu\frac{|t_\nu|^2}{E-\varepsilon_\nu+i0^+},
\end{equation}
where $\varepsilon_\nu$ is the energy of the $\nu$th molecular orbital of the second molecule, and $t_\nu$ is the hopping integral
coupling this orbital with the indicated site of benzene.
The sidegroup introduces Fano antiresonances \cite{clerk01}, which block current through one arm of the annulene, thus lifting the
destructive interference.  Put another way, the second molecule's orbitals
hybridize with those of the annulene, and a state that connects leads 1 and 2 is created in the gap (see Fig.~\ref{suppression}c). 
Shifting $\varepsilon_\nu$ by gating the sidegroup then yields transistor action.

Tunable current suppression occurs over a broad energy range, 
as shown in Fig.~\ref{suppression}b; 
the QuIET functions with any metallic leads 
whose work function lies within the annulene gap.
Fortunately, this is the case for many bulk metals, among them palladium,
iridium, platinum, and gold \cite{marder00}. 
Appropriately doped semiconductor electrodes \cite{piva05} could also be used.

\begin{figure}
\includegraphics[width=\columnwidth,height=.5\columnwidth]{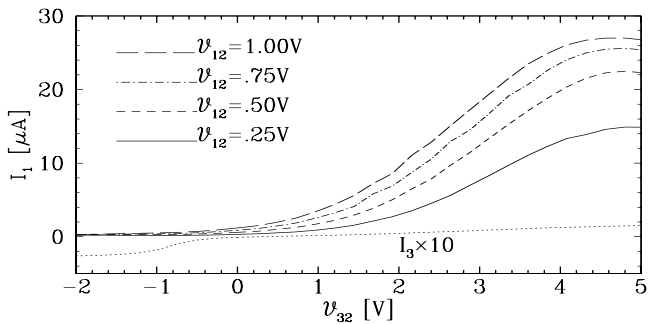}
\caption{$I$--$\mathcal{V}$ characteristic of the QuIET shown in Fig.~\ref{fig:QuIET_3d} 
  at room temperature. The current in lead 1 is shown, where
  $\mathcal{V}_{\alpha\beta}=\mathcal{V}_\alpha-\mathcal{V}_\beta$. Here,
  $\Gamma_1=\Gamma_2=$1eV. 
  $\Gamma_3$ is taken as .0024eV, which allows a small
  current in the third lead, so that the device amplifies current.
  A field-effect device with almost identical $I$--$\mathcal{V}$ can be achieved by taking
  $\Gamma_3=0$. The curve for $I_3$ is for the case of 1.00V bias voltage;
  $I_3$ for other biases look similar.}
\label{IV}
\end{figure}

In Fig.~\ref{IV}, the $I$--$\mathcal{V}$ characteristic of a QuIET based on sulfonated vinyl benzene
is shown, whose molecular structure is given in Fig.~\ref{fig:QuIET_3d}.
The three metallic electrodes were taken as bulk gold, with
$\Gamma_1=\Gamma_2=1\mbox{eV}$, while $\Gamma_3=.0024\mbox{eV}$, so that the coupling
of the third electrode to the alkene sidegroup is primarily electrostatic.
The device characteristic resembles that of a macroscopic transistor.
As the voltage on lead 3 is increased, the antibonding orbital of the alkene sidegroup comes into resonance with the Fermi energies of leads 1 and 2,
leading to a broad peak in the current.
For $\Gamma_{1,2} \gg \Gamma_3\neq 0$, the device amplifies the current
in the third lead (dotted curve), emulating a bipolar junction
transistor. 
For $\Gamma_3=0$, the calculated current $I_1$ is almost identical to that shown in Fig.~\ref{IV}, and 
the device acts like a field effect transistor. 
Alkene chains containing 4 and 6 carbon atoms were also studied, 
yielding devices with characteristics similar to that shown in Fig.~\ref{IV}, with the maximum current $I_1$ shifting to smaller
values of $\mathcal{V}_{32}$ with increasing chain length.  
As evidence that the transistor behavior shown in Fig.~\ref{IV} is due to the tunable interference
mechanism discussed above, we point out that if hopping between the benzene ring
and the alkene sidegroup is set to zero, so that the coupling of the sidegroup to benzene is purely electrostatic,
almost no current flows between leads 1 and 2.

Operation of the QuIET does not depend sensitively 
on the magnitude of the lead-molecule coupling $\bar{\Gamma}=\Gamma_1\Gamma_2/(\Gamma_1+\Gamma_2)$.
The current through the device decreases with decreasing $\bar{\Gamma}$, but aside from that, the 
device characteristic was found to be qualitatively similar when $\bar{\Gamma}$ was varied over one order of magnitude.

\begin{figure}
\includegraphics[keepaspectratio=true,width=\columnwidth]{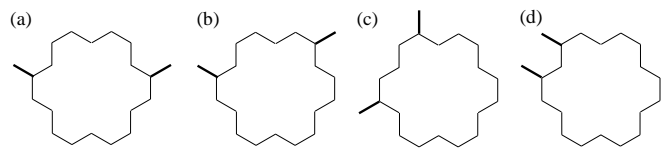}
\caption{Source-drain lead configurations possible in a QuIET
  based on [18]-annulene. The bold lines represent the positioning of the two
  leads. Each of the four arrangements has a different phase difference
  associated with it: (a) $\pi$; (b) $3\pi$; (c) $5\pi$; and (d) $7\pi$.}
\label{fig:18}
\end{figure}

The QuIET mechanism applies to any monocyclic aromatic annulene with leads 1 and 2 positioned so the two most
direct paths have a phase difference of $\pi$. Furthermore,
larger molecules have other possible lead configurations, based on phase differences of $3\pi$, $5\pi$, etc.
Figure \ref{fig:18} shows the lead configurations for a QuIET based on
[18]-annulene.

The position of the third lead affects the degree to
which destructive interference is suppressed.  For benzene, the most effective
location for the third lead is shown in Figs.~\ref{fig:QuIET_3d} and \ref{QuIET}. It may also be
placed at the site immediately between leads 1 and 2, but the transistor effect is somewhat
reduced, since coupling to the charge carriers is less.
The third, three-fold symmetric configuration of leads completely
decouples the third lead from electrons travelling between the first two leads.
For each monocyclic aromatic annulene, 
one three-fold symmetric lead configuration exists, yielding no transistor behavior.

The QuIET's operating mechanism, tunably coherent current suppression,
occurs over a broad energy range within the gap of each monocyclic aromatic annulene;
it is thus a {\em very robust effect, insensitive to moderate fluctuations of the electrical environment of the molecule}.
Although based on an entirely different, quantum mechanical, switching mechanism, the QuIET nonetheless reproduces
the functionality of macroscopic transistors on the scale of a single molecule.

This work was supported by National Science Foundation Grant Nos.\ PHY0210750,
DMR0312028, and DMR0406604, and PHY0244389.

\bibliography{refs}

\end{document}